\global\let\ifmypprint\iffalse 
\def\mypprint{\global\let\ifmypprint\iftrue}
\global\let\iftorefs\iffalse
\def\torefs{\global\let\iftorefs\iftrue}
\global\let\dofloatfig\iffalse
\def\floatthefig{\let\dofloatfig\iftrue}
\mypprint 
\ifmypprint
\documentstyle[preprint,prl,aps,epsf]{revtex}

\draft
\tighten
\floatthefig
\else
  \iftorefs 
    \catcode`\@=11 
    
    \def\figure{\let\@capwidth\columnwidth\@float{figure}}
    \let\endfigure\end@float
    \@namedef{figure*}{\let\@capwidth\textwidth\@dblfloat{figure}}
    \@namedef{endfigure*}{\end@dblfloat}
    \catcode`\@=12 
     \floatthefig 
  \fi
\fi
\input epsf.tex
\begin{document}

\title{The Viscous Nonlinear Dynamics of Twist and Writhe}

\author{Raymond~E. Goldstein,$^{1,2}$ 
Thomas R. Powers,$^{1}$ and Chris H. Wiggins$^{3}$}
\address{$^{1}$Department of Physics and $^{2}$Program in Applied Mathematics}
\address{University of Arizona, Tucson, AZ  85721}
\address{$^{3}$Department of Physics,
Princeton University, Princeton, NJ  08544}

\date{\today}

\maketitle

\begin{abstract}
Exploiting the ``natural" frame of space curves,
we formulate an intrinsic dynamics of twisted elastic
filaments in viscous fluids.
A pair of coupled nonlinear 
equations describing the temporal evolution of the filament's
complex curvature and twist density embodies the dynamic interplay of twist 
and writhe. 
These are used to illustrate a novel nonlinear phenomenon:
{\it geometric untwisting} of open filaments, whereby twisting
strains relax through a transient writhing instability 
without performing axial rotation.
This may explain 
certain experimentally observed motions of fibers of the bacterium
{\it B. subtilis} {[}N.H. Mendelson, {\it et al.},
J. Bacteriol. {\bf 177}, 7060 (1995){]}.

\end{abstract}

\vfil
\eject

As D'arcy Thompson emphasized~\cite{dt},  
nature abounds with examples of
the balance between growth and form.
Thompson was especially interested in 
`dynamic morphology,' the realization of this harmony in organisms.
The balance is particularly manifest in the low Reynolds number regime
of slow motions without inertia, where growth
is generally dominated by dissipation and form is constrained 
by elasticity.  The coupling between these two has been 
termed ``elastohydrodynamics".

\begin{figure}
\epsfxsize=5.0truein
\centerline{\epsffile{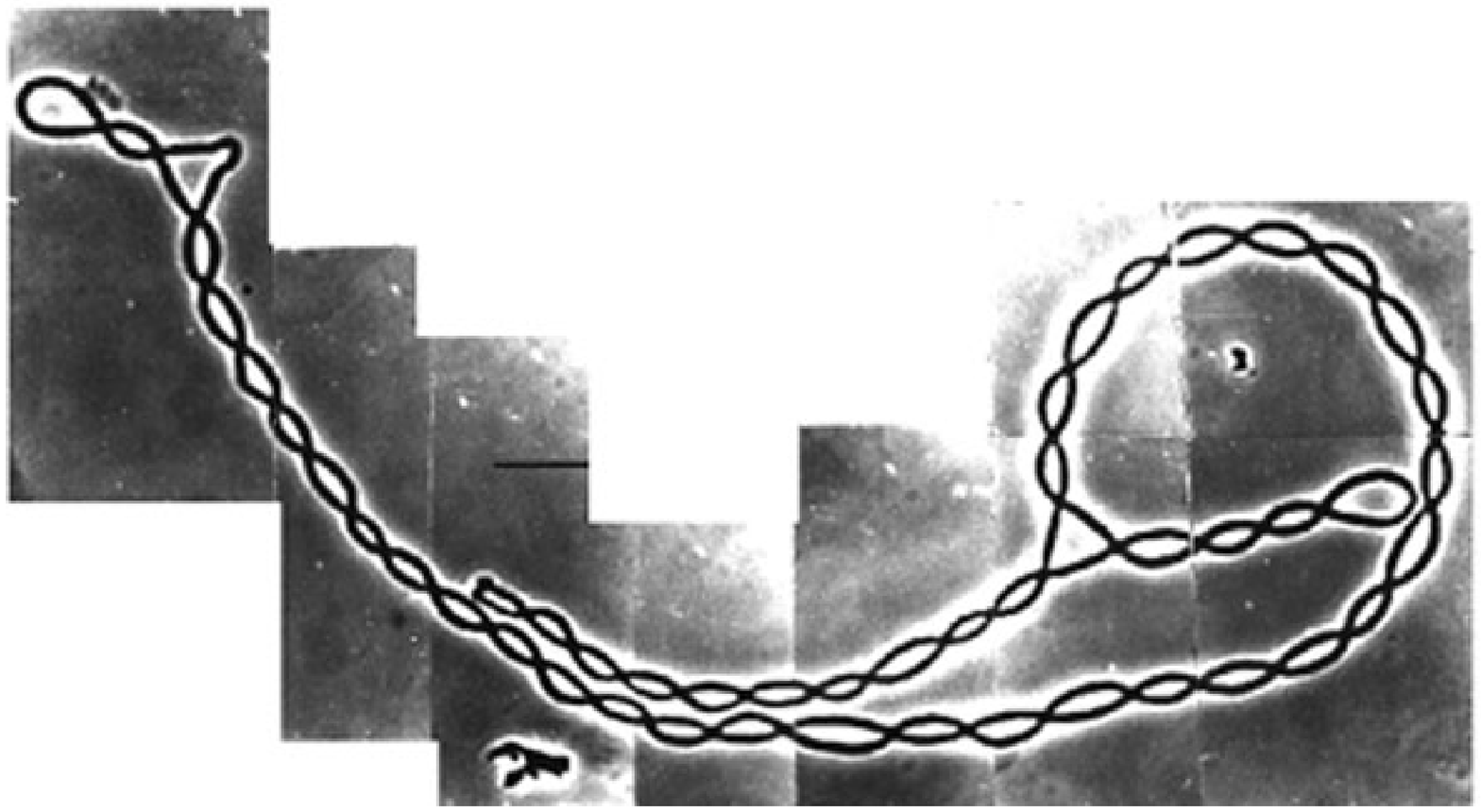}}
\bigskip
\caption{Supercoiled filament of {\it B. subtilis} [3].
Scale bar is $10$ $\mu$m. Image courtesy of N.H. Mendelson.
\label{fig1}}
\end{figure}

Possibly the most common `form' of subcellular dynamics
is the elastic filament:  
rod-like macromolecules perform a multitude of structural
and biological roles. To gain some intuition for
dynamics at this scale, the case of the
elastic filament is thus a fruitful starting point~\cite{ehd}.
As a motivating example, we recall
Mendelson's discovery of shape instabilities exhibited by 
a mutant strain of the bacterium {\it B. subtilis}~\cite{neil1}.
Filaments of cells form as the rod-shaped bacteria fail to separate upon
dividing; beyond a critical length, a filament
buckles,  
twisting around itself to form a 
plectoneme (See Fig.~\ref{fig1}).
It is thought that this instability arises from nonequilibrium twisting
stresses generated in the cell walls through the process of 
growth~\cite{neil1,Tilby}.  As the bacteria continue to grow, this  
process is iterated, leading to a hierarchy 
of supercoils.  The handedness of the coils may be tuned with
temperature or ionic concentrations.
Since the filament ends are generally free and thermal
fluctuations are negligible, this supercoiling
is fundamentally different from that of overtwisted DNA that is closed or
has pinned ends~\cite{alberts}.
As a first step toward understanding these highly nonlinear phenomena, we 
develop here an {\it intrinsic} deterministic formulation of 
twisted elastic filament motion, and apply it 
supercoiling of {\it B. subtilis}.  Various elements of this formulation have
appeared in previous work on 
{\it closed} elastic rods {\it without} dissipation~\cite{klapper}, where 
a {\it dynamic} (rather than energetic~\cite{marko}) coupling between twist
and bend degrees of freedom has been recognized~\cite{klapptab}.
In the inertialess case, a geometric argument for this
coupling has been recently proposed~\cite{kamien}, 
and its qualitative consequences have been explored~\cite{maggs}.  
In presenting a general formulation of filament dynamics,
we explicitly derive this twist-bend coupling and illustrate
the qualitatively new physics of open filaments
in a viscous medium.  

An important feature of our treatment is the parameterization of the filament.
Although the ``Euclidean'' representation based on the position 
vector ${\bf r}(s)$
is convenient for linear stability analysis about simple shapes, it
makes the full nonlinear problem unnecessarily complicated.
As exploited by Kirchhoff~\cite{Kirchhoff}, 
the Euler angles of rigid body motion are better suited for elastic equilibria.  
This parameterization singles out a particular space-fixed reference frame
and thus is partly extrinsic.  An intrinsic representation 
expresses the rate of
change of a frame ({\it e.g.} the material or Frenet-Serret frames) at each 
point of the rod 
in terms of the frame itself~\cite{Landau}, and 
is simplified
by using the ``natural frame'' associated to the centerline.
First used by Darboux~\cite{Darboux} and later 
by Hasimoto in his study of vortex filament 
motion~\cite{hasimoto}, this frame has
been applied to the dynamics of space curves with bending 
but no twisting elasticity~\cite{GnL}.

The configuration of an elastic rod may be described by a 
material orthonormal frame $\{\hat{\bf e}_1,\hat{\bf e}_2,\hat{\bf e}_3\}$
\cite{Landau}: 
we choose $\hat{\bf e}_3$ to point along the tangent,  
$\hat{\bf e}_1$ to point from the centerline to a painted
stripe on the rod surface that connects one end to the other,
and $\hat{\bf e}_2=\hat{\bf e}_3 \times\hat{\bf e}_1$.
The strains $\bbox{\Omega}$ are given by the rate of rotation of 
this frame with arclength $s$, 
$\partial_s\hat{\bf e}_i=\bbox{\Omega}\times\hat{\bf e}_i$.  
To quadratic order in $\bbox{\Omega}$, the elastic 
energy of a homogeneous
isotropic rod is 
\begin{equation}
{\cal E}={A\over2}\int\!{\rm d}s\bigl(\Omega_1^2+\Omega_2^2\bigr)+{C\over2}
\int\!{\rm d}s\Omega_3^2-\int\!{\rm d}s \Lambda,
\label{energystrains}
\end{equation}
where $\Omega_i=\bbox{\Omega}\cdot{\hat{\bf e}}_i,$
$A$ and $C$ are the bend and twist elastic constants respectively,
and $\Lambda(s)$ is a Lagrange multiplier function 
enforcing local inextensibility.

A ``natural'' orthonormal frame~\cite{Darboux} of a given
curve consists of the
unit tangent and two other vectors 
constrained to have zero tangential rotation rate.  Imagine a 
weather vane sliding along the curve and always lying in the 
normal plane~\cite{lanper}.
If the weather vane spins around 
${\hat{\bf e}}_3$ with vanishing initial tangential angular momentum
then it traces out one member of a family of natural frames related by
$s$-independent rotations. 
Such a frame is constructed simply by untwisting the material frame,
{\it i.e.,}\ rotating ${\hat{\bf e}}_1$ and 
${\hat{\bf e}}_2$ around ${\hat{\bf e}}_3$ by minus the accumulated
twist angle $\vartheta(s)\equiv\int^s\!{\rm d}s^\prime\Omega_3$.
Since $\bbox{\epsilon}^{*}\cdot\partial_s\bbox{\epsilon}=0$, the vector 
$\bbox{\epsilon}\equiv({\hat{\bf e}}_1+i{\hat{\bf e}}_2)
\exp i\vartheta$ does not twist around ${\hat{\bf e}}_3$. 
  
The rotation rate of this
frame is given by the complex curvature $\psi$: 
$\partial_s\bbox{\epsilon}=-\psi{\hat{\bf e}}_3$ and $\partial_s 
{\hat{\bf e}}_3={\rm Re}(\psi\bbox{\epsilon}^{*})$, 
where
\begin{equation}
\psi
=\bigl(-i\Omega_1+\Omega_2\bigr)\exp i\int^s\!\!{\rm d}s^\prime\Omega_3.
\label{psi}
\end{equation}
Some elementary shapes have simple
$\psi$-representations: the straight line has $\psi=0$,
the circle $\psi=a$, and the helix
$\psi=a{\rm e}^{ips}$ ($a,p$ are real constants).  Likewise, certain
geometric properties of curves are easily stated in terms of 
$\psi$:  a curve
lies in a plane if and only if $({\rm Re}\psi(s), {\rm Im}\psi(s))$
lies on a line through the origin in the complex $\psi$-plane,
and a curve lies on a sphere if and only if 
$({\rm Re}\psi(s), {\rm Im}\psi(s))$
lies on a line not through the origin~\cite{Bishop}.

We compute from (\ref{energystrains}) the elastic forces and moments
per unit length using the principle of virtual work.  If ${\bf F}$ is
the force acting on a cross-section, the force per unit
length ${\bf F}_s$
is found by varying the position of the centerline without
rotating any element of the rod about the tangent direction:  
$\delta{\cal E}=-\int{\rm d}s{\bf F}_s\cdot\delta{\bf r}(s)$, with 
the rotation angle
$\delta\chi\equiv{\hat{\bf e}}_2\cdot\delta{\hat{\bf e}}_1=0$. 
Likewise, if ${\bf M}$ is the moment acting
on a cross section, the tangential component of the 
moment per unit length ${\bf M}_s$ is found by varying the
orientations of the elements of the rods without moving the centerline:
$\delta{\cal E}=-\int{\rm d}s{\bf M}_s\cdot{\hat{\bf e}}_3\delta\chi$, with
$\delta{\bf r}=0$.  Using $\delta {\bf r}_s\equiv(\delta{\bf r})_s$, 
$\delta\hat{\bf e}_a=(\delta\chi)\epsilon_{ab}{\hat{\bf e}}_b-
(\hat{\bf e}_a\cdot\delta{\bf r}_s)\hat{\bf e}_3$ for $a,b=1,2$ and 
$\delta\hat{\bf e}_3=\delta{\bf r}_s-(\hat{\bf e}_3\cdot\delta
{\bf r}_s){\hat{\bf e}}_3$, we find 
\begin{eqnarray}
\delta\psi&=&-2\psi{\hat{\bf e}}_3\cdot\delta{\bf r}_s
+\bbox{\epsilon}\cdot\delta{\bf r}_{ss}
-i\psi{\rm Im}\int^s\!\!{\rm d}s^\prime{}\psi\bbox{\epsilon}
\cdot\delta{\bf r}_s,\label{psivar}\\
\delta\Omega_3&=&\delta\chi_s-\Omega_3\hat{\bf e}_3\cdot\delta{\bf r}_s+
(\Omega_1{\hat{\bf e}}_1+\Omega_2{\hat{\bf e}}_2)\cdot\delta{\bf r}_s.
\label{psiomvar}
\end{eqnarray}
Note that keeping $\delta\chi=0$ during the variation does not imply
that the twist rate $\Omega\equiv\Omega_3$ stays fixed!  (For brevity
we drop the subscript on $\Omega_3$ below.)

Thus the force per unit length is
$-\delta{\cal E}/\delta{\bf r}=-{\bf e}_3\Lambda_s+
{\rm Re}\left(\bbox{\epsilon}{{{\cal F}_{\perp}}}\right)$,
where 
\begin{equation}
{{{\cal F}_{\perp}}}=-A\bigl(\psi_{ss}+{1\over2}|\psi|^2\psi\bigr)+
iC\bigl(\Omega\psi\bigr)_s
-\Lambda\psi,
\label{gam}
\end{equation}
and a term $C\Omega^2/2$ has been absorbed into $\Lambda$.  
Separating (\ref{gam}) into its real and imaginary
parts yields the classic results
of Love \cite{love}. 
Note that the force density is
invariant under $\psi\mapsto e^{i\phi}\psi$,
$\bbox{\epsilon}\mapsto e^{i\phi}\bbox{\epsilon}$ for constant $\phi$.
The tangential moment per unit length is $-\delta{\cal E}/\delta\chi(s)=
C\Omega_s$.  The conditions of zero force and moment at free ends
imply the boundary conditions $\psi=\psi_s=\Lambda=\Omega=0$.

The dynamics of $\psi$ and $\Omega$ follow immediately
from~(\ref{psiomvar}) by 
choosing $\delta{\bf r}={\bf r}_t\delta t$ and 
$\delta\chi=\chi_t\delta t$.
Their form is completely dictated by geometry.  
(That for $\psi_t$ is equivalent to earlier results
\cite{GnL}; the undamped analogue of the $\Omega_t$ equation
appears in~\cite{klapptab}.) The 
right-hand side of~(\ref{psiomvar}) 
expresses a fact easily demonstrated with the nearest rubber tube;
there are two distinct
ways of changing the twist density---(i) nonuniform axial rotations of the 
rod, and (ii) nonplanar bending motions.  

To complete the equations of motion, 
we use the simplest Rouse dynamics
in which the only effect of the
surrounding fluid is to provide a local isotropic drag~\cite{Doi}
proportional to the local velocity:  
${\bf f}_{\rm drag}=\zeta{\bf r}_t$.
Likewise, we take the tangential
moment per unit length to be $m_{\rm drag}=\zeta_{\rm r}\chi_t.$
Since at zero Reynolds number the elastic forces and moments 
must exactly balance the drag
forces and moments, the complex curvature and twist obey
\begin{eqnarray}
\zeta\psi_t&=&\bigl(\partial_s^2+|\psi|^2\bigr){{{\cal F}_{\perp}}}+i\psi
{\rm Im}\int^s\!\!{\rm d}s^\prime{{{\cal F}_{\perp}}}^*\psi_s
-\Lambda_s\psi_s 
\label{eoms_final1}\\
\zeta_{\rm r}\Omega_t&=&C\Omega_{ss}
+(\zeta_{\rm r}/\zeta){\rm Im}{{{\cal F}_{\perp}}}_s^*\psi~.
\label{eoms_final}
\end{eqnarray}
Local inextensibility, 
$\partial_t{\rm d}s=0$,
is assured by demanding 
$\Lambda_{ss}={\rm Re}(\psi^*{{{\cal F}_{\perp}}}).$
Eq. (\ref{eoms_final1}) is covariant and (\ref{eoms_final}) invariant
under $s$-independent
rotations of the natural basis about the tangent:  $\psi\mapsto e^{i\phi}\psi$.
Since the elastic energy (\ref{energystrains}) is achiral,
the dynamics are covariant under mirror reflections:
$\psi\mapsto\psi^*$, $\Omega\mapsto-\Omega$.

The derivation of the above dynamics relies solely
on geometry and mechanics.  However, for closed curves,
we can reconsider (\ref{eoms_final}) in light of the
celebrated relation ${{\cal L}k}={{\cal T}w}+{{\cal W}r}$ \cite{fuller}. 
Here ${{\cal L}k}$ is the linking number of the two curves 
traced out by the heads of the vectors ${\hat{\bf e}}_1$ and ${\hat{\bf e}}_2$
(assuming a continuous material frame),
${{\cal T}w}\equiv\int{\rm d}s\Omega/(2\pi)$, and the writhe ${{\cal W}r}$ is
an integer plus $A/(2\pi)$, where $A$ is the signed area enclosed by the 
curve swept out on the unit sphere by ${\hat{\bf e}}_3$~\cite{fuller2}.  
For a locally inextensible filament, the twist dynamics (\ref{eoms_final}) 
can be interpreted as a 
local link conservation law under the assumption that smooth filament motions
will not change the integer contribution to the writhe\cite{kamien}.  
The time derivative of the 
density $wr$ of the noninteger contribution to the writhe is
$\partial_t wr={\hat{\bf e}}_3\cdot\partial_t{\hat{\bf e}}_3
\times\partial_s{\hat{\bf e}}_3/(2\pi)=
-{\rm Im}({{{\cal F}_{\perp}}}_s^*\psi)/(2\pi\zeta)$.  
Therefore, the twist dynamics has the form $\rho_t+j_s=0$, with the ``link
density'' $\rho=\Omega/(2\pi)+wr$ and link current
$j=-(C/\zeta_{\rm r})\Omega_s$.  Of course, since the link is not a
single integral of a local density there is no true link density;
however, {\it changes} in link can be written as a single integral
over a density~\cite{fuller2}.
While ${{\cal L}k}$ and ${{\cal W}r}$ are conventionally only defined 
for closed filaments, 
the equations of motion (\ref{eoms_final1}) and (\ref{eoms_final}) 
show that the tradeoff between twist and writhe still has meaning for 
open filaments.

With the coupled evolution equations for $\psi$ and
$\Omega$ we turn to linear and nonlinear aspects of
supercoiling instabilities.  The simplicity of the $\psi$-formulation
for open filaments is seen first in the stability
of a straight rod with a {\it uniform} twist density 
(temporarily neglecting the boundary conditions on $\Omega$).  
The linearization of (\ref{eoms_final1}) is
\begin{equation}
\zeta\psi_t=-A\psi_{ssss}+iC\Omega\psi_{sss}~.
\label{psilin}
\end{equation}
The eigenfunctions $W_n$ of $-A\partial_s^4+iC\Omega\partial_s^3$ 
have eigenvalues
$\sigma_n=-Ak^4_n+C\Omega k_n^3$ (and associated growth rates $\sigma_n/\zeta$) 
which are real since the operator is Hermitian.
If $\Omega=0$, the modes are planar shapes with
$W_n= c_n^{(1)}\sin(k_ns)+c_n^{(2)}\cos(k_ns)+
c_n^{(3)}\sinh(k_ns)+c_n^{(4)}\cosh(k_ns),$
with $\cos(k_nL)\cosh(k_nL)=1$~\cite{Landau}. 
Since the $W_n(s)$ satisfy the boundary conditions and 
form a complete set, the shape of {\it any} free 
elastica in three dimensions, whether tightly wound helices or
plectonemes, is expressible as
\begin{equation}
\psi(s,t)=\sum_{n=1}^{\infty} C_n(t)W_n(s)~.
\label{psi_rep}
\end{equation}
The time evolution is a nonlinear dynamical system in
$C_n(t)\equiv\langle W_n\vert\psi(t)\rangle$, where the $W_n$ are normalized.

\begin{figure}
\epsfxsize=5.0truein
\centerline{\epsffile{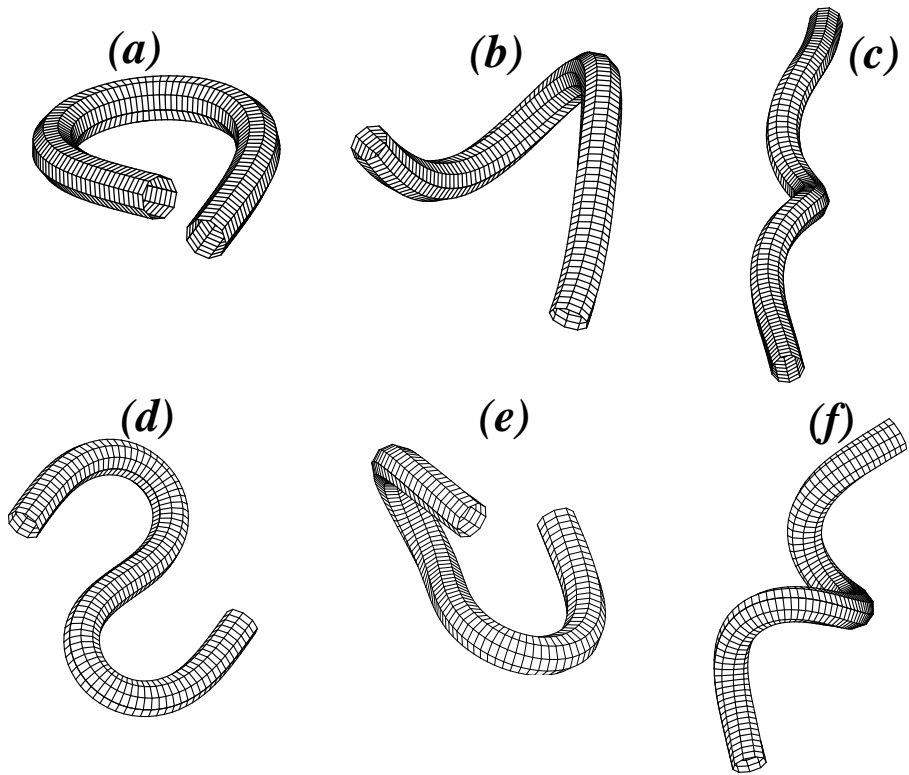}}
\smallskip
\caption{Filament shapes at various twist densities. (a)-(c): $\psi=10W_1$
at $\Omega/\Omega_c^{(1)}=0.22,1.0,2.22$; (d)-(f): $\psi=8W_2$ at
$\Omega/\Omega_c^{(2)}=0.13,0.58,1.29$.  
\label{fig2}}
\end{figure}

The linear stability analysis of a free rod with uniform twist
is identical to that of a clamped rod
with uniform twist.  Using the complex Monge representation 
$\xi(z)=x(z)+iy(z)$, where ${\bf r}(z)=(x(z),y(z),z)$, yields
$\zeta \xi_t=-A\xi_{zzzz}+iC\Omega\xi_{zzz}$ with $\xi_z=\xi_{zz}=0$
at the clamped ends.  Note however that the Monge representation is
incapable of describing plectonemes. 
The critical condition for the first centerline
instability is $\Omega_c\approx8.98 A/CL$~\cite{Landau}.
When the twist is non-zero, each
eigenfunction is a superposition of four helices
with appropriate spatial damping:
$W_n(s)=\sum_{\mu=1}^4 c_n^{(\mu)}\exp(ik_n^{(\mu)}s)$, where for each $n$ the 
four $k_n^{(\mu)}$ are in general complex.  Modes with negative
eigenvalue $\sigma_n$ have three helices of one handedness 
and one of the other;
otherwise, all have the same handedness.
Shapes calculated from the first two modes are shown in Fig.~\ref{fig2}.  

The writhe-tracking term in (\ref{eoms_final})
is explicitly nonlinear; when the diffusion constant $C/\zeta_{\rm r}=0$,
twist only changes when the filament writhes.   
In assessing the relative importance of 
twist diffusion and writhe-tracking we note that while
$\zeta_r$ in (\ref{eoms_final}) was interpreted as 
a rotational drag coefficient, in the context of 
bacterial filaments it is not 
clear that this dissipative
mechanism dominates others, {\it e.g.,} rearrangement
of the polymer network inside the 
cell walls.  If the shape instabilities are generated by
nonequilibrium twist, then twist must not relax on time scales shorter
than the observed buckling time (tens of minutes \cite{initiation}).
Using the purely hydrodynamic twist diffusion constant $C/\zeta_{\rm r}$ 
\cite{maggs,vistwist}
(with $C\equiv10^{-10}$ erg-cm \cite{Thwaites} and 
$\zeta_r\equiv10^{-10}$ erg-sec/cm 
for a one micron diameter slender rod in water), 
and the typical filament length $L\sim 100$ $\mu$m for buckling, 
the time scale $\tau_D=L^2\zeta/C$ for
twist to diffuse out the ends of the filament
is only $10^{-4}$ sec, eliminating hydrodynamic
drag as a candidate mechanism. 
This and other observations ({\it e.g.} of slow stress 
relaxation \cite{Thwaites}) 
strongly suggest a dominant
dissipative mechanism {\it within} the filament, 
characterized by a very much larger $\zeta_r$.   
Therefore, we consider the limit of negligible diffusion constant $C/\zeta_r$, 
where twist changes only by bending rather than by axial
rotation of the filament, a relaxation mechanism we term 
``geometric untwisting."

\begin{figure}
\epsfxsize=5.0truein
\centerline{\epsffile{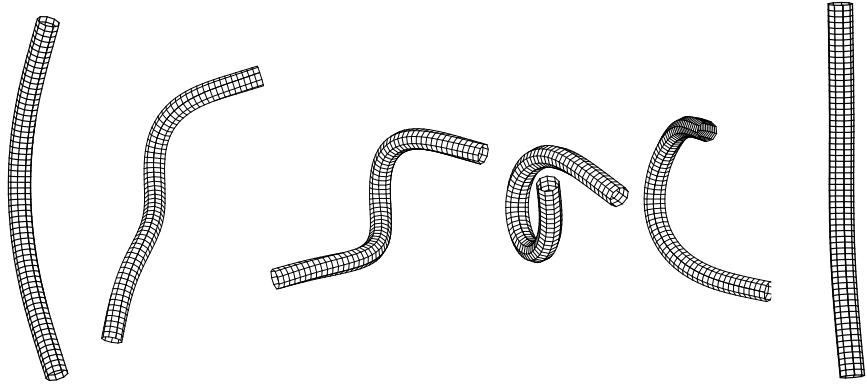}}
\smallskip
\caption{Geometric untwisting.  Filament shapes are calculated from
numerical integration of (\ref{eoms_final1}) and (\ref{eoms_final}) 
with $C/A=1$.
Time proceeds from left to right.   
\label{fig3}}
\end{figure}

\begin{figure}
\epsfxsize=5.0truein
\centerline{\epsffile{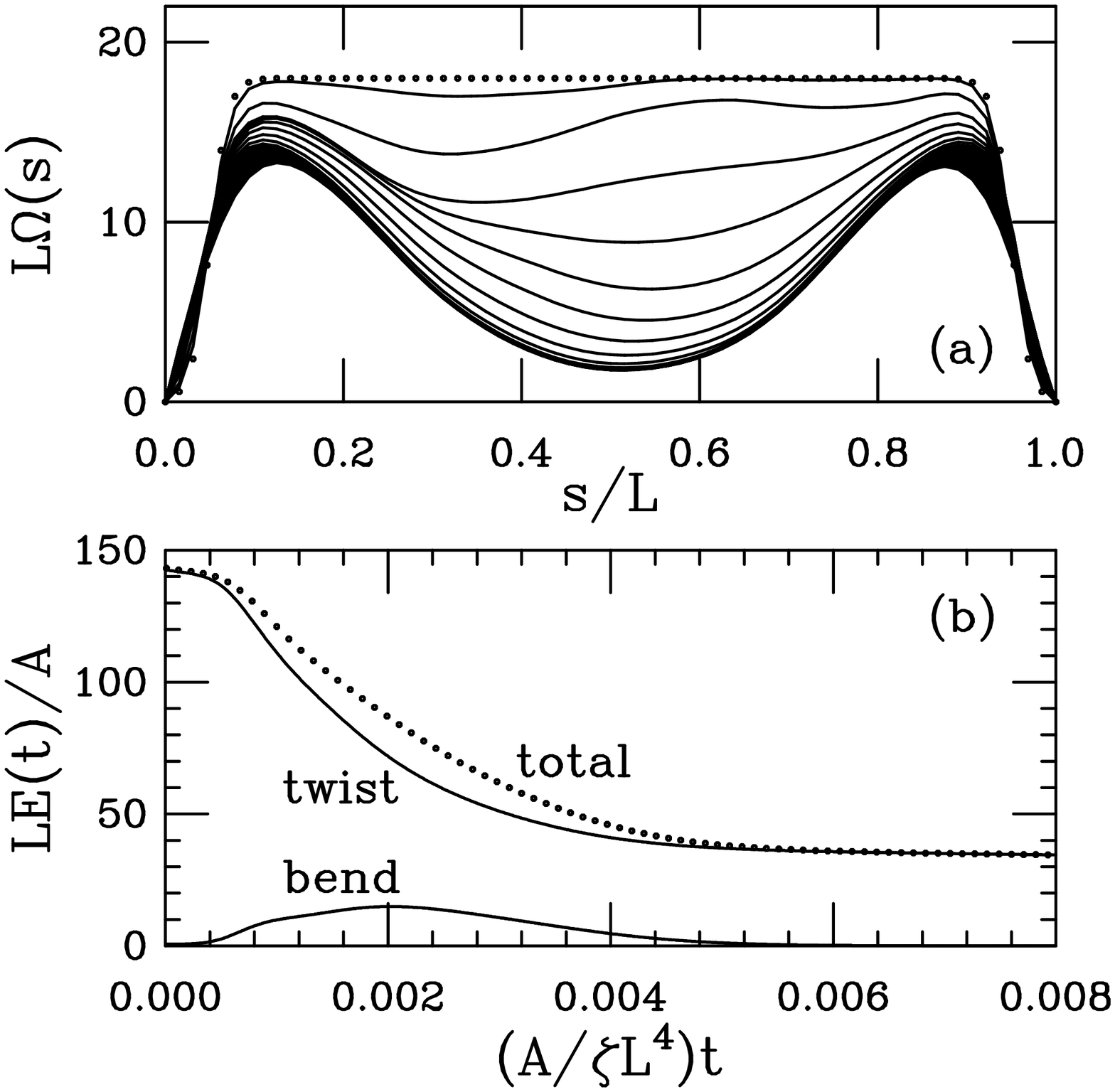}}
\smallskip
\caption{Details of geometric untwisting.  
(a) Twist density evolution, with dotted
line showing initial profile. Asymmetry in $\Omega$ arises from that of the
initial shape.  (b) Time evolution of twist, bend, and total
elastic energy.
\label{fig4}}
\end{figure}

Geometric untwisting is illustrated in Fig. \ref{fig3}, obtained by numerical
integration \cite{numerics} of (\ref{eoms_final1})
and (\ref{eoms_final}).  An additional nonlocal contact force in
${{{\cal F}_{\perp}}}$ prevents self-crossing.
The initial condition is
a nearly straight filament with twist density of the form
shown in Fig.~\ref{fig4}a.  This mesa-like profile reflects
uniform local twist 
production during growth and diffusive boundary layers at the
ends~\cite{wpg}.  
Motion begins as chiral buckling; as
the curvature increases, twist is converted to writhe, and
$\Omega$ decreases in the interior (Fig. \ref{fig4}a). 
When twist has sufficiently relaxed and writhe is of order unity, 
the filament relaxes back to
the straight ground state as link flows out the ends.
In this overdamped dynamics the total elastic energy is monotonically
decreasing, despite the transient increase in the bending energy
during loop formation (Fig. \ref{fig4}b).

The sequence in Fig. \ref{fig3} is quite similar to that
exhibited by short filaments of {\it B. subtilis} \cite{initiation}:
after beginning the transition
to a plectoneme via a looping instability, the ends touch only
transiently, the loop snaps open, and the filament straightens.  
Plectoneme formation occurs after several
aborted looping events, presumably as the twist stresses continually
increase with filament extension.

The results presented here proceed from the robust to the specific:
from kinematics dictated solely by geometry,
through viscous dynamics governed by elasticity, and united in
a description of the onset of bacterial supercoiling.
This particular union will help unravel the puzzles
surrounding growth-induced iterated writhing instabilities 
of bacterial fibers \cite{wpg} and elucidate possible twist-induced
instabilities in related systems, such as thermally fluctuating 
nucleic acids \cite{Wang}. 

We are indebted to Neil Mendelson and Charles Wolgemuth for ongoing
collaborations, and thank A. Goriely, G. Huber, R. Kamien, J. Kessler,
and M. Tabor for important discussions.  This work was supported by 
NSF Presidential Faculty Fellowship DMR96-96257 (REG).

\begin{thebibliography}{000}

\bibitem{dt}D'arcy W. Thompson, {\it On Growth and Form} (The University 
Press, Cambridge, 1942).

\bibitem{ehd} D. Riveline, C.H. Wiggins, R.E.
Goldstein, A. Ott, Phys. Rev. E. {\bf 56}, 1330 (1997);
C.H. Wiggins, D. Riveline, R.E. Goldstein, and A. Ott, 
Biophys. J. {\bf 74}, 1043 (1998).

\bibitem{neil1} N.H. Mendelson, Proc. Natl. Acad. Sci. USA {\bf 73}, 1740 
(1976); Microbiol. Rev. {\bf 46}, 341 (1982).

\bibitem{Tilby} M.J. Tilby, Nature {\bf 266}, 450 (1977).

\bibitem{alberts} B. Alberts, D. Bray, J. Lewis, M. Raff, K. Roberts, and
J.D. Watson, {  Molecular Biology of the Cell}, 3rd ed. (Garland Publishing,
Inc., New York, 1994).

\bibitem{klapper} I. Klapper, J. Comp. Phys. {\bf 125}, 325 (1996);
A. Goriely and M. Tabor, Phys. Rev. Lett. {\bf 77}, 3537 (1996).

\bibitem{marko} J.F. Marko and E.D. Siggia, Macromolecules {\bf 27}, 981
(1994).

\bibitem{klapptab} I. Klapper and M. Tabor, J. Phys. A: Math. Gen.
{\bf 27}, 4919 (1994).

\bibitem{kamien} R.D. Kamien, Eur. Phys. J. B {\bf 1}, 1 (1998).

\bibitem{maggs} A.C. Maggs, preprint, 1997:  cond-mat/9704054.

\bibitem{Kirchhoff} G. Kirchhoff, J.F. Math (Crelle) {\bf 50}, 285 (1859).

\bibitem{Landau} See {\it e.g.} L.D. Landau and E.M. Lifshitz, 
{\it Theory of Elasticity}, 3rd ed. (Pergamon Press, Oxford, 1986).

\bibitem{Darboux} G. Darboux, {\it Le\c cons sur la Th\' eorie Gen\' erale
des Surfaces} (Gauthier-Villars, Paris, 1915), Vol. I, p. 22. 

\bibitem{hasimoto} H. Hasimoto, J. Fluid Mech. {\bf 51}, 477 (1972).

\bibitem{GnL} R.E. Goldstein and S.A. Langer, Phys. Rev. Lett. {\bf
75}, 1094 (1995); K. Nakayama, H. Segur, and M. Wadati, Phys. Rev.
Lett. {\bf 69}, 2603 (1992).

\bibitem{lanper} Joel Langer, private communication.  Langer defines
a natural frame to be one with constant tangential component of 
angular velocity. 

\bibitem{Bishop} R.L. Bishop, Amer. Math. Monthly {\bf 82}, 246 (1975).

\bibitem{love}  A.E.H. Love, {\it A Treatise on the Mathematical 
Theory of Elasticity}, 4th ed. (Cambridge, London, 1965).  See also  Y. 
Shi and J.E. Hearst, J. Chem. Phys. {\bf 101}, 5186 (1994).


\bibitem{Doi} M. Doi and S.F. Edwards, {\it The Theory of Polymer Dynamics} 
(Oxford University Press, New York, 1986).

\bibitem{fuller} G. C\u alug\u areanu, Rev. Math. Pures Appl. 
{\bf 4}, 5 (1959); J.H. White, Am. J. Math. {\bf 91}, 693 (1969); 
F.B. Fuller, Proc. Nat. Acad. Sci. USA {\bf 68}, 815 (1971).

\bibitem{fuller2} F.B. Fuller, Proc. Natl. Acad. Sci. USA {\bf 75},
3557 (1978).


\bibitem{initiation} N.H. Mendelson, J.J. Thwaites, J.O. Kessler, 
and C. Li, J. Bacteriol. {\bf 177}, 7060 (1995).

\bibitem{vistwist} M.D. Barkley and B.H. Zimm, J. Chem. Phys. {\bf 70},
2991 (1979); J.M. Schurr, Chemical Physics {\bf 84}, 71 (1984).  

\bibitem{Thwaites} J.J. Thwaites and N.H. Mendelson, Int. J. Biol. Macromol.
{\bf 11}, 201 (1989).

\bibitem{numerics} Computations employ the representation 
(\ref{psi_rep}); stiffness constraints are removed as in 
T.Y. Hou, J.S. Lowengrub, and M.J. Shelley, J. Comp. Phys. {\bf 114}, 
312 (1994).

\bibitem{wpg} C.H. Wiggins {\it et al.}, in preparation.

\bibitem{Wang} L.F. Liu, J.C. Wang, Proc. Natl. Acad. Sci. (USA) {\bf 84}, 
7024 (1987); C. Levinthal and H.R. Crane, {\it ibid.} {\bf 42}, 436 (1956).

\end{thebibliography}
\end{document}
\end